\documentclass{iau}
\usepackage{graphicx}
\title[JD 11.~~Selection of Massive Clusters]
{Galaxy and Mass Assembly (GAMA): \\ Selection of the Most Massive Clusters.}

\author[H{\'e}ctor J. Ibarra-Medel, Maritza Lara-L{\'o}pez, \& Omar L{\'o}pez-Cruz .]
{H{\'e}ctor J. Ibarra-Medel$^{1,2}$, Maritza Lara-L{\'o}pez$^3$, Omar L{\'o}pez-Cruz$^1$ \and  the GAMA Team.}

\affiliation{$^{1}$National Institute of Astrophysics, Optics and Electronic, \\ 
Luis Enrique Erro No 1, Tonantzintla, Puebla, M{\'e}xico \\ email: {\tt omarlx@inaoep.mx} \\[\affilskip]
$^2$ email: {\tt ibarram@inaoep.mx} \\[\affilskip]
$^3$Institute of Astronomy, National Autonomous University of M{\'e}xico, \\ Box
70-264, M{\'exico} City, M{\'e}xico \\ email: {\tt maritza@astro.unam.mx}}

\pubyear{2014}
\volume{308}  
\pagerange{119--126}
\jname{The Zeldovich Universe Genesis and Growth of the Cosmic Web}
\editors{Rien van de Weygaert, Sergei Shandarin, Enn Saar \& Jaan Einasto, eds.}
\begin{document}

\maketitle

\begin{abstract}
We have  developed a galaxy cluster finding technique based on the Delaunay Tessellation Field Estimator (DTFE) combined with caustic analysis. Our method allows us to recover clusters of galaxies  within the mass range of $10^{12}$ to $10^{16}\ \mathcal{M}_{\odot}$. We have found a total of 113 galaxy clusters in the Galaxy and Mass Assembly survey (GAMA). In the corresponding mass range, the density  of clusters found in this work is comparable to the density traced by  clusters selected by the thermal Sunyaev Zel'dovich Effect; however, we are able to cover a wider mass range.  We present the analysis of the two-point correlation function for our cluster sample. 
\keywords{methods: data analysis, surveys, galaxies: clusters: general, cosmology: miscellaneous, cosmology: observations}
\end{abstract}

\firstsection 
\section{Introduction}

The GAMA survey (http://www.gama-survey.org/) is a multi-wavelength spectroscopic survey that covers $\sim360$ deg$^2$, which includes $\sim400,000$ galaxy redshifts down to a magnitude limit of $r_{AB}=19.8$ (\cite[Driver et al.  2011]{DR11}). We chose three stripes within GAMA that cover $\sim144$ deg$^2$ with $\sim110,000$ galaxy spectra. These three equatorial sky stripes are centred at 9h, 12h and 14.5h (\cite[Driver et al. 2011]{DR11}).

We have  implemented a new  cluster finding technique to find overdensities and estimate cluster masses, simultaneously.   We find number galaxy  overdensities by using an adaptive method based on the Delaunay Tessellation Field Estimator  (\cite[DTFE, Schaap \& van de Weygaert 2000,  Platen 2009]{SW00,platen}), mass estimation is done using caustic analysis (\cite[e.g., Serra et al. 2011,  Alpaslan et al. 2012]{S11,A12}).  We use this method to detect clusters of galaxies  within the mass range of $10^{12}$ to $10^{16}\ \mathcal{M}_{\odot}$, up to  $z=0.3$.

\begin{figure}[t]
\begin{center}
\includegraphics[width=2.72in]{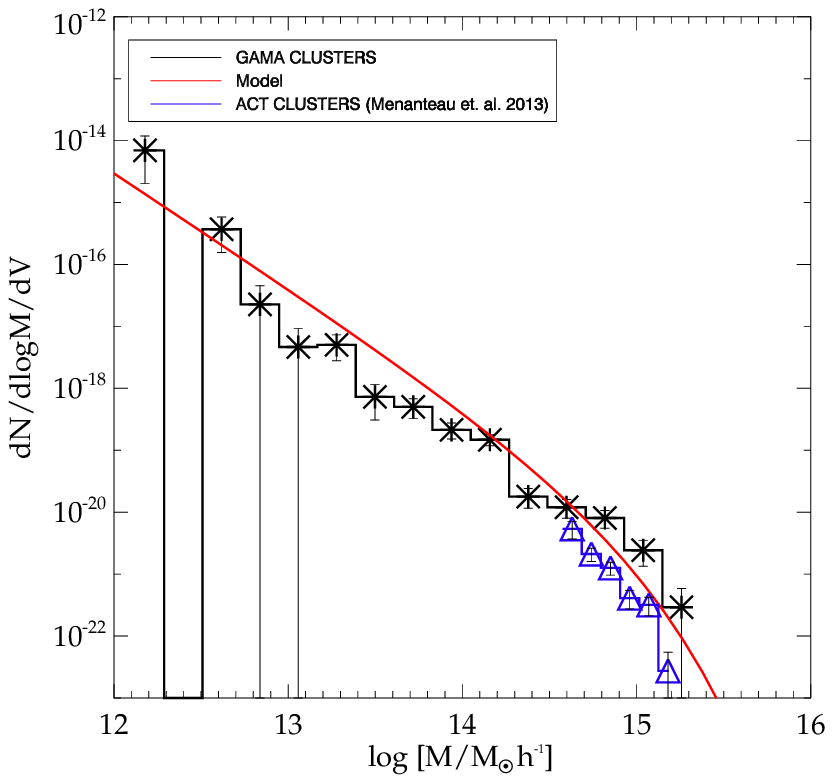} 
\includegraphics[width=2.55in]{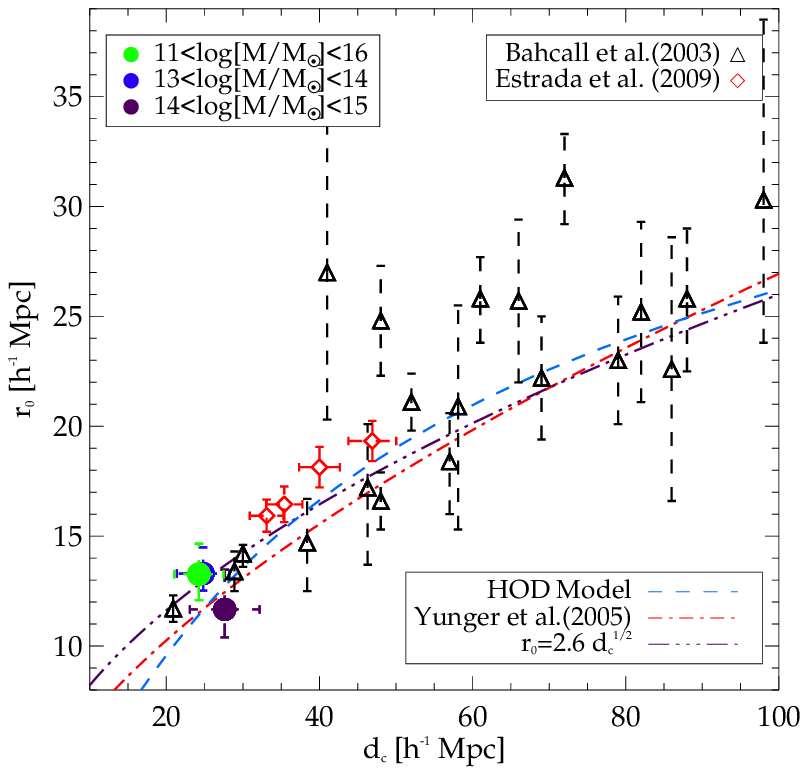}
\caption{ The left panel present the comoving density traced by the cluster found in this study (indicated by filled stars), we have compare our results those of Menanteau et al, (2013) generated from a sample of massive clusters selected by the Sunyaev-Zel'dovich effect (open triangles), the continuous dashed line is a simple halo occupation model (HOD).  In the right panel we present a compilation of previous results on the characteristic scale for the two-point cluster correlation as a function of cluster separation, the filled big  dots represents the results for our sample.  We find agreement with previous studies and models}\label{fig1}
\end{center}
\end{figure}

\section{Overview and Results}

We have found 113 cluster within GAMA.  For this sample we have  estimated positions,  cluster redshifts, velocity dispersions, cluster sizes,   and  cluster integrated luminosity.  Our algorithm  has been tested using the GAMA mock catalogs (\cite[Robotham et al. 2011]{R11}). The  calculation  the cluster luminosities have been generated by using the individual  cluster-galaxy luminosity functions (LF) corrected for completeness. We have evaluated the  cluster selection function  by the application of a simple halo occupation distribution (HOD) model. We want to stress  that the density of clusters found by mass selection methods (e.g., the Atacama Cosmology Telescope (ACT), \cite[Menanteau et al. 2013]{M13}) is comparable to one found in this work; however, we have covered a larger mass range by more than  three orders of magnitude. In addition, we have generated the two-point  correlation  for clusters of galaxies for our sample. We find broad agreement previous observations and predictions  (\cite[Estrada et al. 2009]{Es09}).  We have generated the mass-to-light ratio (M/L) for the clusters  and BCGs in our sample, we find that a single power law $\mathcal{L}\propto\mathcal{M}^{\eta}$ can describe . We found  $\eta=0.6-1$ for clusters and $\eta_{BGC}=0.1-0.4$ for BCGs. These relations agree with the results of \cite[Lin et al.(2004)]{YM04a} and \cite[Lin \& Mohr(2004)]{YM04b}.  

The sample found in this study can be used for further studies in galaxy evolution and its relation with environment. We have shown that optical surveys such as GAMA can be used to select cluster by mass.  A sample of cluster selected by our method can be used to traced baryon acoustic oscillations using in a survey in which galaxies are selected in the same fashion as GAMA but covering a larger volume.

\end{document}